\documentclass[runningheads]{llncs}
\usepackage[T1]{fontenc}
\usepackage{graphicx}
\usepackage{booktabs}
\usepackage{color}
\usepackage{multirow}
\usepackage{siunitx}

\newcommand\blfootnote[1]{%
  \begingroup
  \renewcommand\thefootnote{}\footnote{#1}%
  \addtocounter{footnote}{-1}%
  \endgroup
}

\begin{document}
\title{Power-Capping Metric Evaluation for Improving Energy Efficiency in HPC Applications}
% alternatives:
% Decision-maker metrics for improving Energy Efficieny of HPC Applications with Power Capping
% Decision-maker metrics for improving GPU Energy Efficieny of HPC Applications with Power Capping

% Metric Evaluation for HPC Applications with Power Capping to achieve Energy Efficieny
% Power Capping Metric Evaluation for Energy-Efficient HPC Applications
% Power Capping Decision-maker Metric Evaluation for Energy-Efficient HPC Applications

\titlerunning{Power-Capping Metric Evaluation for Improving Energy Efficiency}
\author{
Maria Patrou\inst{1}\orcidID{0000-0003-3975-4638} \and 
Thomas Wang\inst{2}\orcidID{0009-0001-2491-5323} \and \\
Wael Elwasif\inst{1}\orcidID{0000-0003-0554-1036} \and   
Markus Eisenbach\inst{1}\orcidID{0000-0001-8805-8327} \and  \\
Ross Miller\inst{1}\orcidID{0000-0002-2179-495X} \and
William Godoy\inst{1}\orcidID{0000-0002-2590-5178} \and
Oscar Hernandez\inst{1}\orcidID{0000-0002-5380-6951}
}
\authorrunning{M. Patrou et al.}
\institute{
Oak Ridge National Laboratory, Oak Ridge, Tennessee, USA\\
\email{\{patroum,elwasifwr,eisenbachm,rgmiller,godoywf,oscar\}@ornl.gov} \and
Camas High School, Camas, Washington, USA\\
\email{twangapples@gmail.com}
}
\maketitle              % typeset the header of the contribution
\begin{abstract}
    \begin{sloppypar}
With high-performance computing systems now running at exascale, optimizing power-scaling management and resource utilization has become more critical than ever. This paper explores runtime power-capping optimizations that leverage integrated CPU-GPU power management on architectures like the NVIDIA GH200 superchip. We evaluate energy-performance metrics that account for simultaneous CPU and GPU power-capping effects by using two complementary approaches: speedup-energy-delay and a Euclidean distance-based multi-objective optimization method. By targeting a mostly compute-bound exascale science application, the Locally Self-Consistent Multiple Scattering (LSMS), we explore challenging scenarios to identify potential opportunities for energy savings in exascale applications, and we recognize that even modest reductions in energy consumption can have significant overall impacts. Our results highlight how GPU task-specific dynamic power-cap adjustments combined with integrated CPU-GPU power steering can improve the energy utilization of certain GPU tasks, thereby laying the groundwork for future adaptive optimization strategies.
\end{sloppypar}

    \keywords{Power Capping, Energy Efficiency, HPC, GH200, Performance Metrics, Automatic Power Steering, Exascale Applications, LSMS}
\end{abstract}

\blfootnote{
Notice: This manuscript has been authored by UT-Battelle LLC under contract DE-AC05-00OR22725 with the US 
Department of Energy (DOE). The US government retains and the publisher, by accepting the article for 
publication, acknowledges that the US government retains a nonexclusive, paid-up, irrevocable, worldwide 
license to publish or reproduce the published form of this manuscript, or allow others to do so, for US 
government purposes. DOE will provide public access to these results of federally sponsored research in 
accordance with the DOE Public Access Plan (https://www.energy.gov/doe-public-access-plan).
}

\section{Introduction}
GPUs have become crucial for exascale computing because of their high performance per watt compared with traditional CPUs, thus enabling significant improvements in application performance across diverse computational tasks. As a result, the high-performance computing (HPC) community has widely adopted GPUs and developed specialized programming models and numerical methods to maximize the benefits of GPU acceleration.

However, the slowdown of Moore's law has caused processors to lag behind the increasing compute and power demands of scientific application workloads. Thus, optimizing power and energy efficiency on HPC systems remains a critical and ongoing challenge. Although previous research has focused on optimizing application workloads and enabling facility-level improvements, more work is needed to fine-tune applications through dynamic tuning~\cite{10.1145/3126908.3126945}, mixed-precision methods, and computational batching to utilize accelerators more effectively~\cite{haidar2019investigating}. Additionally, emerging hardware capabilities present new opportunities. Tightly integrated CPU-GPU superchips, such as the NVIDIA GH200, incorporate an automatic power-steering system that dynamically reallocates power between the CPU and GPU. This system can be used in conjunction with power capping to further improve energy efficiency. 
These new capabilities can now be exploited to improve the application's energy efficiency by applying finer-grained power capping based on the type of computations of the GPU tasks. 
Additionally, meaningful power-capping metrics play a critical role in guiding users to select appropriate optimization goals and tailor power-management strategies to individual application GPU tasks, thereby enabling users to define customized power-management settings for each application.

To this end, we took an initial step toward adaptive power-capping strategies by evaluating metrics to guide power-cap selection for a highly optimized exascale application. Even modest improvements in energy efficiency can yield substantial benefits at exascale, thus highlighting the relevance of this exploration. We used the NVIDIA GH200 Grace Hopper superchip and applied power-capping limits to control the chip's total (CPU-GPU) power consumption. We used automatic power steering to allow CPUs and GPUs to draw power within the set limits based on the application demands. For our study, we targeted key computational tasks, GPU kernels and GPU compute idle phases (where the CPU is executing work), of the Locally Self-Consistent Multiple
Scattering (LSMS) exascale application (see Section~\ref{appkernel}) and identified the regions most impacted by the power-capping feature. We examined ways in which GPU task-specific power tuning can improve energy efficiency on compute- and memory-bound kernels within the self-consistent field (SCF) iterations. In our analysis, we evaluated and compared different decision-making metrics to identify suitable GH200 superchip power settings based on the execution characteristics of each GPU task. The speedup-energy-delay and Euclidean distance of normalized energy/runtime metrics provide complementary insights into power-capping effects and assist in finding optimized power settings for different GPU tasks executed during different application phases. We compared the results provided by each methodology and present their impacts on energy and runtime performance in ideal scenarios to identify the appropriate test cases for each one.
Through our findings, we characterized scientific GPU tasks, identified methods for selecting appropriate power-cap settings, and contributed to power-capping tuning strategies that achieve greater energy efficiency.

\section{Background}
Research has shown that power capping improves energy efficiency, reduces hardware failures~\cite{zhao2023sustainable}, and extends system lifespan. 
Haidar et al.~\cite{haidar2019investigating} and Krzywaniak et al.~\cite{krzywaniak2022gpu} demonstrated how composite metrics, particularly energy-delay product (EDP), help manage GPU power limits and balance energy savings and performance. Abdulsalam et al.~\cite{abdulsalam2015gpsup} introduced the Greenup, Powerup, and Speedup (GPS-UP) framework to categorize optimizations by their impact on execution time, power, and energy efficiency. Fine-grained power-capping strategies, such as the Global Extensible Open Power Manager~\cite{10.1007/978-3-319-58667-0_21}, EE-HPC~\cite{10820573}, and READEX~\cite{gerndt2016readex}, have proven more effective than static limits. Our approach is similar to fine-grained dynamic power optimizations but is driven by the exploration of effective optimization metrics.

Modern GPUs and CPUs independently regulate power via capping~\cite{10.1007/978-3-030-43222-5_11}, but new-generation architectures, such as NVIDIA's GH200 superchip, regulate both CPU and GPU simultaneously. The GH200 incorporates power capping within its automatic power-steering system and dynamically reallocates power between the CPU and the GPU based on their usage. It initially allocates power to the CPU, thus transferring any unused capacity to the GPU, and enforces limits through Dynamic Voltage and Frequency Scaling (DVFS). When GPU power usage nears a power limit, the system reduces GPU clock speeds. This reduction affects compute-intensive workloads more significantly than memory-bound tasks. Other approaches for CPU-GPU power management with heuristic or machine learning methods have been explored by Saba et al.~\cite{saba2023ml} and Azimi et al.~\cite{azimi2023powercoord}, who dynamically allocate power between CPU and GPU. Our approach leverages the automatic power steering of the GH200 when setting a given power cap, thereby dynamically optimizing power allocation per application GPU task to improve both energy efficiency and performance.

\subsection{Application}
\label{appkernel}
The LSMS exascale application~\cite{EISENBACH20172,10.1145/3581784.3607089} is a density functional theory code designed to compute the electronic properties of materials---particularly metals, alloys, and nanostructures. Instead of diagonalizing the Hamiltonian, LSMS utilizes the Korringa-Kohn-Rostoker method, which efficiently determines electronic interactions by computing Green’s function. This approach allows LSMS to scale efficiently and support simulations beyond 100,000 atoms. 
The most computationally demanding tasks within LSMS involve dense-matrix operations, including matrix multiplications and block inversions, which contribute significantly to execution time. The code iteratively updates the electron density and potential through an SCF loop. Written in a combination of C\texttt{++} and Fortran, LSMS achieves high scalability by leveraging the Message Passing Interface (MPI) for atom-level parallelization, OpenMP on CPUs for energy-level calculations, and CUDA/HIP to accelerate dense linear algebra operations on GPUs. This hybrid programming approach enables LSMS to maintain near-linear scaling on large-scale exascale supercomputing platforms~\cite{10.1145/3581784.3607089}.

\section{Methodology}
%\section{Methodology}
Our approach to optimize LSMS involves leveraging power capping to identify optimal maximum power settings that improve both energy efficiency and runtime performance. In this study, we systematically investigated the impact of varying power-capping levels on both fine-grained computational GPU tasks and overall application performance. We measured and analyzed execution time (seconds), energy consumption (joules), and power (watts) at the superchip level for the entire application, and we recorded the CPU- and GPU-specific power for each GPU task within the application run. These measurements were taken while applying incremental chip-level power constraints on the GH200 superchip, ranging from 200\,W to the maximum power limit of 1,000\,W (default). We executed the entire application at every superchip power limit and collected the data for every GPU task and idle period.

We collected the power data by using Score-P~\cite{10.1007/978-3-642-31476-6_7} in combination with a custom Performance Application Programming Interface (PAPI) component~\cite{10.1145/3703001.3724383} designed to read power information from the GH200 monitoring infrastructure via the Linux \textit{/sys/class/hwmon} interface. Two Score-P plug-ins were employed: one used this PAPI component to monitor superchip and CPU power, whereas the other interfaced with the NVIDIA Management Library (NVML) to measure GPU power. Each plug-in ran on a dedicated thread per node and sampled power data every 5\,ms. We calculated the total energy and runtime values in each run and aggregated the measurements together per GPU task. We performed three runs per power-cap setting and calculated the average values (total energy and runtime) per GPU task and power-cap setting. The runtime overhead per run for the power measurements was 1.3\%. By evaluating these detailed power profiles and performance metrics across various power-cap settings, we calculated specific decision-making metrics to reveal the optimal power-limit configurations. This comprehensive analysis enables informed decision-making to maximize energy efficiency while minimizing application runtime performance degradation. Although the specific optimal power settings can vary across applications, the methodology and metrics can be adapted and used in every application/kernel of interest.

To evaluate our metrics and GPU task-specific power-capping strategies, we ran experiments by using the LSMS application (Section~\ref{appkernel}) 
%with Score-P 8.4 instrumentation (OpenMP, CUDA, and ArmPL library wrappers) 
on the Wombat testbed's NVIDIA GH200 nodes. Each node includes a 72-core Arm Neoverse-V2 CPU (3.52 GHz), 480\,GB error-correction code (ECC) LPDDR5X memory, and a GH200 GPU with 96\,GB ECC HBM3 memory. The software environment used Red Hat Enterprise Linux 9.4, NVIDIA driver 560.35.03, CUDA 12.6, an NVHPC 24.9 compiler suite, and GNU Compiler Collection 11.4.1. For our experimental runs, we configured LSMS to run with a single MPI rank, one OpenMP thread (\texttt{OMP\_NUM\_THREADS=1}), and one GPU per rank. These settings resemble those of an exascale MPI rank, thus ensuring precise, analyzable, and reproducible conditions. For the application input, we adjusted the \textit{iron platinum} benchmark case\footnote{\texttt{lmax=7}; \texttt{rLIZ=18}; \texttt{rsteps={89.5, 91.5, 93.2, 99.9}}; \texttt{atom="Fe"}; and \texttt{Z=26}} to achieve high GPU utilization. 

Subsequent subsections detail the kernel selection process, the power-capping levels applied, and the analysis methods used to evaluate and identify the most energy-efficient configurations across different computational scenarios within LSMS. 

\subsection{Baseline Run of LSMS}
\label{baseline}

\begin{figure}[htbp]
    \centering
    \includegraphics[clip, trim=8 10 4 0,width=\textwidth]{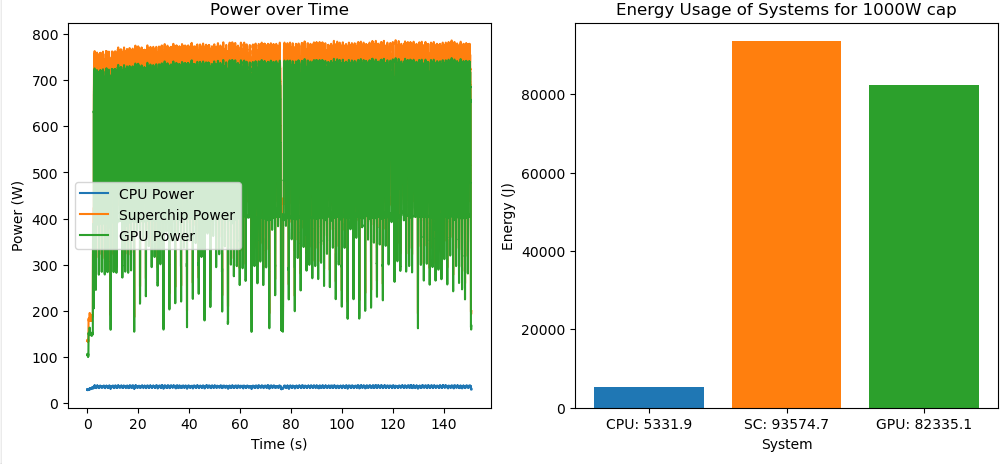}
    \caption{Power and total energy consumed by LSMS on the GH200 superchip (SC), including CPU and GPU components, with the default maximum power setting of 1,000\,W (no power capping).}
    \label{fig:powertime}
\end{figure}
We executed LSMS on the GH200 superchip by using the default power limit of 1,000\,W (no power capping) as a baseline scenario. We measured kernel execution times along with power and total energy consumption for both CPU and GPU.
The total runtime was 150.79\,s, and the total energy consumption was 93,574.7\,J. Fig.~\ref{fig:powertime} shows the power traces for the superchip, GPU, and CPU during the execution of LSMS and shows the cumulative energy usage for each component: 82,335.1\,J for the GPU and 5,331.9\,J for the CPU. The GPU predominantly drove both power consumption and total energy usage for this application. The trace distinctly captured two SCF iterations, and observable power drops between iterations marked the transition of computations between the GPU and the CPU. Superchip power data was collected through \texttt{hwmon}. We subtracted CPU power from superchip power~\cite{nvidia_power_thermals}, thus yielding higher-resolution GPU power data points suitable for our fine-grained, GPU task-level analysis.

Because most power consumption occurs in the GPU, we focused on analyzing GPU kernels. Table~\ref{tab:kernel_summary_en} provides details of the GPU kernel execution with the default power setting, including total runtime, invocation counts, cumulative energy, and average power consumption. GPU tasks are listed in order of total energy consumption, thereby providing a clear prioritization for subsequent power-capping optimizations.
\begin{table}[htbp]
    \small
    \centering
    \caption{Measurements at the default power setting (1,000\,W, no power capping) per GPU kernel and GPU compute idle time.}
    %Summary of GPU kernel executions on the GH200 superchip at the default power setting (1000 watts, no power capping), showing total runtime, number of invocations, total energy consumption, and average power.
    \label{tab:kernel_summary_en}
    \resizebox{\textwidth}{!}{\begin{tabular}{p{3.8cm}|r|r|r|r}
        \toprule
        \textbf{Task} & \textbf{Total Time (s)} & \textbf{\# Calls} & \textbf{Total Energy (J)} & \textbf{Avg. Power (W)} \\
        \midrule
        sm90\_gemm\_ts64x64x32 & 77.89 & 21,632 & 35,361.83 & 454.02 \\
        buildKKRMatrix & 34.90 & 128 & 12,867.73 & 368.74  \\
        sm90\_gemm\_ts32x32x32 & 8.03 & 94,208 & 4,076.98 & 507.51  \\
        getrf\_pivot(1) & 4.07 & 16,384 & 2,694.54 & 662.05  \\
        getrf\_pivot(2) & 4.07 & 30,720 & 2,670.36 & 656.11  \\
        trsm\_left\_kernel & 3.57 & 150,272 & 2,328.26 & 651.57  \\
        getrf\_pivot(3) & 1.82 & 8,192 & 1,146.70 & 630.06  \\
        gpu compute idle & 8.83 & 601,345 & 2,425.49 & 274.80  \\
        \bottomrule
    \end{tabular}}
\end{table}
The kernel \texttt{sm90\_gemm\_ts64x64x32} consumed the most energy overall (35,361.83\,J). This amount far exceeded the second highest energy consumption (12,867.73\,J), which came from the kernel \texttt{buildKKRMatrix}. However, invocation counts differed notably: \texttt{sm90\_gemm\_ts64x64x32} was invoked 21,632 times, whereas \texttt{buildKKRMatrix} was invoked only 128 times. This finding highlights the importance of analyzing each kernel's individual contribution to the execution characteristics to inform effective power-management strategies. We also profiled the GPU idle periods (\texttt{gpu compute idle}). These are phases in the application in which the CPU performs computations while the GPU remains idle. The phases occur mostly between SCF iterations. Identifying the GPU compute idle phases can inform design strategies for reducing GPU power with minimal impact on GPU runtime performance and steering the computational power to CPUs for efficient CPU runtime performance.

Below is a brief description of tasks
that contribute significantly to both energy consumption and runtime:

\begin{itemize}
    \item \texttt{sm90\_gemm\_ts64x64x32}: Compute-intensive cuBLAS zgemm operation; typically compute-bound %sm90\_xmma\_gemm\_tilesize64x64x32
    \item \texttt{buildKKRMatrix}: Matrix construction primarily involving memory operations; memory-bound. %buildKKRMatrixMultiplyKernelCuda
    \item \texttt{sm90\_gemm\_ts32x32x32}: Another compute-intensive cuBLAS zgemm kernel; compute-bound %sm90\_xmma\_gemm\_tilesize32x32x32
    \item \texttt{getrf\_pivot} kernels: Lower-upper factorization with pivoting; constrained by random memory accesses. We investigated three function invocations with different parameters.
    \item \texttt{trsm\_left\_kernel}: Triangular matrix solve (BLAS); typically memory-bound due to data access patterns
    \item \texttt{gpu compute idle}: Periods in which computations only occur on the CPU
\end{itemize}

\begin{sloppypar}
These tasks represent distinct computational phases within LSMS: \texttt{buildKKRMatrix} constructs the multiple-scattering matrix, while the other kernels mainly solve dense linear equations. Identifying each kernel's characteristics is essential for fine-tuning GPU task-specific power caps to optimize energy efficiency and minimize the runtime performance impact.
\end{sloppypar}

\subsection{Metrics}
Traditional HPC optimizations prioritize execution time speedup. However, this performance alone does not guarantee energy efficiency because higher performance can come at the cost of increased power consumption. Achieving optimal energy efficiency requires exploring an optimization space that balances runtime and energy consumption. To evaluate this balance, we employed two complementary metrics: speedup-energy-delay~\cite{nvidia_energy_efficiency} and Euclidean distance of normalized energy/runtime~\cite{9969367}, each capturing different aspects of energy-performance trade-offs. We investigated the most appropriate power-cap settings in terms of energy efficiency and runtime performance for fine-grained computational regions (GPU tasks). These metrics help us to clearly quantify the impact of the underlying hardware power behavior as we scale the power-cap settings. Comparing the most energy-efficient solutions as defined by each metric helps us understand how a decision-making approach affects energy/runtime optimizations and determine the suitability of each metric. 

\subsubsection{Speedup-Energy-Delay Metric.}

To better evaluate the trade-offs between execution time and energy consumption, we employed a variation of the energy-delay product (EDP), known as the \textit{speedup-energy-delay metric}. This metric provides a way to evaluate power optimizations by incorporating both performance improvements and energy efficiency into a single formula.
This metric~\cite{nvidia_energy_efficiency} is particularly useful for comparing different optimizations on HPC systems because it highlights the balance between reducing execution time and minimizing energy consumption. It is defined as follows:

\[
SPEEDUP_{energy-delay} = \frac{\left[\frac{runtime_1}{runtime_n}\right]}{\left[\frac{energy_n}{energy_1}\right]} = \frac{[runtime_1 \ast energy_1]}{[runtime_n \ast energy_n]},
\]

\noindent where
\begin{itemize}
    \item $runtime_1$ and $energy_1$ refer to the baseline runtime and energy consumption, respectively, and
    \item $runtime_n$ and $energy_n$ represent the optimized runtime and energy consumption, respectively.
\end{itemize}

The purpose of this metric is to minimize the product of runtime and energy consumption for the optimized version of the code, as expressed by

\[
\min \ [runtime_n \ast energy_n].
\]

This formulation allows us to evaluate optimizations in a way that reflects both performance and energy trade-offs, thereby providing a useful basis for comparison with other metrics.

\subsubsection{Euclidean Distance of Normalized Energy/Runtime Metric.}
This metric was previously used to identify the CPU frequency that offers the most efficient combination of energy consumption and runtime performance of web requests by applying the CPU DVFS technique~\cite{9969367}. It is based on the Global Criterion method for multi-objective optimization using Euclidean distance, which yields a Pareto-optimal solution, where no objective can be improved without compromising the other.~\cite{9610368}. We adapted the approach to evaluate GPU tasks at different power-cap settings while focusing on GPU accelerators. 

First, we normalized the GPU energy $n_{energy_{ki}}$ for every GPU task $k$ and power-cap setting $i$ by using the feature scaling normalization method (min-max); see the formula below.  
We collected the total energy values for every GPU task at every available power-cap setting (200--1,000\,W). Focusing on each GPU task, we found the minimum ($emin_k$) and maximum ($emax_k$) energy among these nine power-cap settings. The normalization process yielded values in the interval $(0, 1)$.
\[
n_{energy_{ki}} =\frac{[e_{ki} - emin_k]}{[emax_k -emin_k]},
\]

where
\begin{itemize}
    \item $n_{energy_{ki}}$ refers to the normalized energy for a GPU task $k$ at a specific power-cap setting $i$, 
    \item $e_{ki}$ refers to the energy measured for a specific GPU task $k$ and power-cap setting $i$,
    \item $emin_k$ refers to the minimum energy measured for a specific GPU task $k$ across all power-cap settings, and
    \item $emax_k$ refers to the maximum energy measured for a specific GPU task  $k$ across all power-cap settings.
\end{itemize}
Equivalently, we calculated the normalized runtime (\(n_{runtime_{ki}}=\frac{[r_{ki} - rmin_k]}{[rmax_k -rmin_k]} \)).
Because energy and runtime are expressed at different scales, the normalization process ensures that we can include both metrics together in equal terms in subsequent calculations. 

Second, we calculated the Euclidean distance of the normalized energy and normalized runtime for each GPU task $k$ at every power-cap setting $i$:
\[
distance_{ki} = \sqrt{n_{energy_{ki}}^2 + n_{runtime_{ki}}^2}.
\]
The purpose of this metric is to find the global minimum distance for each GPU task 
\[ min[distance_{ki}]  \] 

and indicate the power-cap setting that corresponds to it. It should indicate that with the selected power-cap setting, the scaled energy consumption cannot be decreased any further without increasing the scaled runtime performance~\cite{9610368}, thereby revealing the optimal power-cap setting for efficiency.

\section{Experimental Evaluation}
%\subsection{Metrics Comparison}
\begin{figure}[htbp]
    \centering
        \includegraphics[clip, trim=10 8 2 10,width=\textwidth]{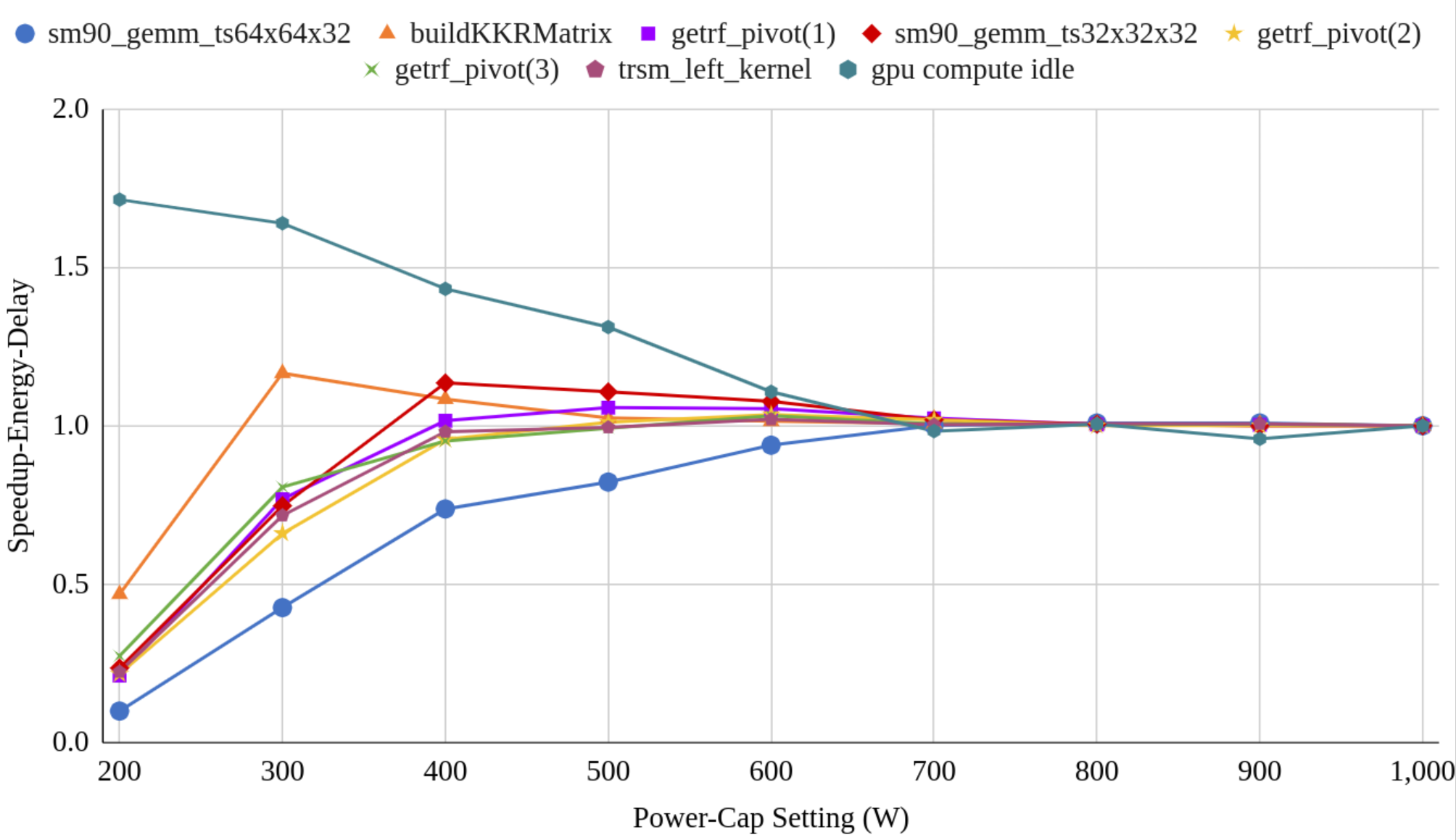}
    \caption{Speedup-energy-delay per GPU task and power-cap setting. Higher is better.}
    \label{fig:speedup}
\end{figure}

\begin{sloppypar}
Fig.~\ref{fig:speedup} shows the speedup-energy-delay metric across different power-cap settings for each previously described GPU task. The plot illustrates different behaviors depending on the type of GPU task and computational characteristics. For most GPU tasks, an initial significant increase in speedup-energy-delay is observed when increasing power-cap settings from the lowest level of 200\,W, generally peaking between 300\,W and 600\,W with the exception of \texttt{sm90\_gemm\_ts64x64x32}, which is a compute-bound GPU task that reaches its peak at 900\,W. After the peak, the metric generally slightly declines and stabilizes. This behavior indicates that further increases in the power cap do not yield any benefits.
Kernels such as \texttt{trsm\_left\_kernel}, \texttt{getrf\_pivot(1)}, \texttt{getrf\_pivot(2)}, and \texttt{getrf\_pivot(3)} show similar speedup-energy-delay trends and cluster around optimal power-cap settings of 600\,W; \texttt{getrf\_pivot(1)} reaches its optimal setting at 500\,W. The \texttt{buildKKRMatrix} kernel, which is memory bound, reaches optimal speedup-energy-delay at a lower power cap of 300\,W and stabilizes after 400\,W.
\end{sloppypar}

The \texttt{gpu compute idle} phase closely follows the trend observed in the energy metric: it reaches its optimal speedup-energy-delay at the lowest power-cap setting (200\,W) with a speedup value of 1.71, after which the values progressively decrease with increasing power settings. This graph highlights the importance of considering GPU task-specific characteristics when determining the best power-cap settings to achieve energy efficiency without substantial performance trade-offs.

\begin{figure}[htbp]
    \centering
        \includegraphics[clip, trim=4 8 4 10,width=\textwidth]{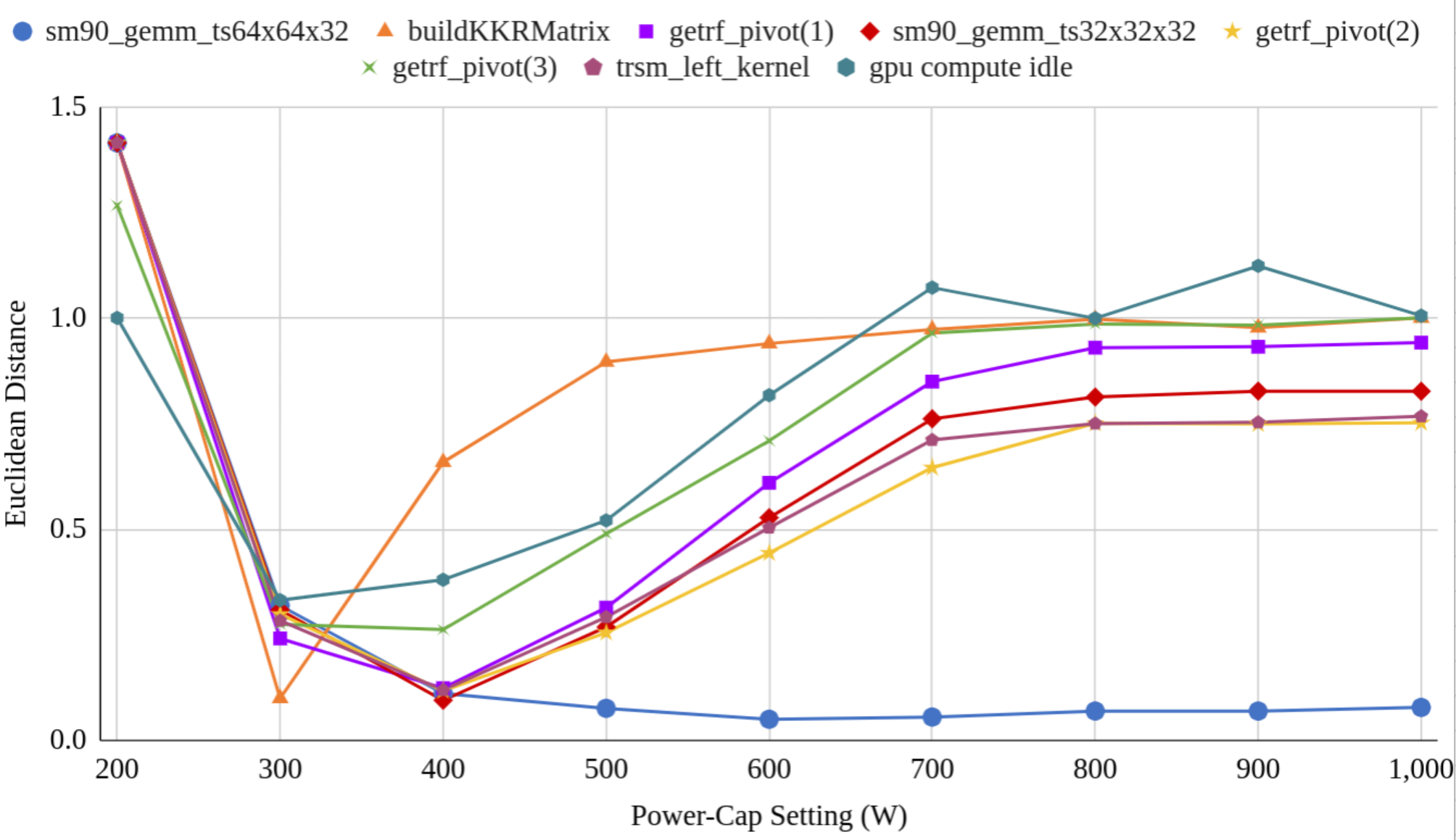} % before eu_distance_plot_total.pdf
    \caption{Euclidean distance of normalized energy/runtime per GPU task and power-cap setting. Lower is better.}
    \label{fig:distance}
\end{figure}

\begin{sloppypar}
Fig.~\ref{fig:distance} shows the Euclidean distance of GPU tasks for each power-cap setting.  
Most of the GPU tasks show the maximum distance at 200\,W, so this is the least suitable setting according to the current metric. In some cases, the calculated distances exceed 1. This power setting increases execution time and produces the slowest runtimes; it also increases energy consumption and yields some of the highest energy usage. As we scale up the power cap, the GPU tasks' distances decrease and reach a minimum between 300\,W and 600\,W. Most of the GPU tasks reach their minimum distances at 400\,W (five GPU tasks), and then their distances increase. At 700\,W, their distances begin to stabilize. Thereafter, any power-cap setting does not substantially alter the distance values.
The kernels \texttt{trsm\_left\_kernel}, \texttt{getrf\_pivot(1)}, 
\texttt{getrf\_pivot(2)}, and \texttt{sm90\_gemm\_ts32x32x32} follow this pattern. Their trends form the tightest cluster among all GPU tasks. The pattern presents similarities with the CPU-driven requests described in previous work~\cite{9610368} and reveals a compute-intensive characteristic. The \texttt{getrf\_pivot(2)} kernel follows a similar pattern.
The \texttt{buildKKRMatrix} kernel demonstrates the minimum at 300\,W, and the distance remains largely unchanged for power-cap increases after 600\,W.
In contrast, the distance for the \texttt{sm90\_gemm\_ts64x64x32} kernel decreases at 600\,W and then stabilizes at higher power-cap settings. Finally, the \texttt{gpu compute idle} phase reaches a minimum distance at 300\,W, where it exhibits the second lowest energy consumption, and a maximum distance at 900\,W, where the highest energy consumption is observed. The total energy consumption for this case increases as the power setting increases. Execution time is highest at the lowest power setting, and it fluctuates as the power increases. This fluctuation is similarly reflected in the distance metric. With the current measurements and metric, the memory-bound \texttt{buildKKRMatrix} and \texttt{gpu compute idle} phase exhibit the same optimal power setting.
\end{sloppypar}

\begin{table}[htbp]
    \scriptsize
    \centering
    \caption{Runtime increase (\%) and Energy reduction (\%): default power compared with the most suitable power setting identified by speedup-energy-delay (SED) and Euclidean distance (ED) per GPU task.}
    \label{tab:comp_all}
    \resizebox{\textwidth}{!}{\begin{tabular}{p{3.8cm}|p{1cm}|p{1cm}|r|r|r|r}
        \hline
        \multirow{2}{*}{\textbf{GPU task}} & \multicolumn{2}{r|}{\textbf{Power Cap (W)}} & \multicolumn{2}{r|}{\textbf{Energy (\%)}} & \multicolumn{2}{c}{\textbf{Runtime (\%)}} \\
        & \textbf{SED} & \textbf{ED} & \textbf{SED} & \textbf{ED} & \textbf{SED} & \textbf{ED}\\
        %\textbf{Kernel} & \textbf{SED} & \textbf{ED} & \textbf{SED Energy (\%)} & \textbf{ED Energy (\%)} & \textbf{SED Runtime (\%)} & \textbf{ED Runtime (\%)} \\
        \hline
        sm90\_gemm\_ts64x64x32 & 900 & 600 & 0.85 & 3.42 & 0.00 & 10.3 \\
        buildKKRMatrix & 300 & 300 & 22.92 & 22.92 & 11.30 & 11.30 \\
        sm90\_gemm\_ts32x32x32 & 400 & 400 & 31.40 & 31.40 & 28.42 & 28.42 \\
        getrf\_pivot(1) & 500 & 400 & 20.61 & 28.50 & 19.16 & 37.67 \\
        getrf\_pivot(2) & 600 & 400 & 10.05 & 24.48 & 7.37 & 38.41 \\
        trsm\_left\_kernel & 600 & 400 & 9.02 & 25.53 & 7.74 & 36.85 \\
        getrf\_pivot(3) & 600 & 400 & 9.10 & 24.15 & 6.59 & 38.46 \\
        gpu compute idle & 200 & 300 & 46.58 & 39.69 & 9.25 & 1.17 \\
        \bottomrule
    \end{tabular}}
\end{table}

For a comparison between the most suitable power-cap settings (as defined by each metric) and the system’s default setting (1,000\,W), Table~\ref{tab:comp_all} presents the percent change in energy and runtime for each GPU task. The energy change is calculated as \( \frac{energy_i-energy_{1000}}{energy_{1000}}*100 \), in which the negative sign ($-$) is omitted, indicating energy reduction. The runtime change 
(or increase) is calculated as \( \frac{runtime_i-runtime_{1000}}{runtime_{1000}}*100 \). In a few cases, both metrics define the same optimal power-cap setting. However, in most cases, they point to nearby settings that can affect energy savings and runtime performance quite differently.

Through a comparison of these two metrics, we identified significant opportunities for fine-grained energy optimization, especially notable in GPU tasks such as \texttt{buildKKRMatrix} and during GPU idle states. This optimization yields considerable energy savings with a small performance slowdown cost. GPU idle states are critical targets for power-capping optimization because they may occur frequently in HPC. Additionally, analyzing GPU task invocation frequency against energy consumption shows that even infrequently invoked GPU tasks can be good candidates for GPU task-specific power optimizations.

By aggregating the observed percentage values with speedup-energy-delay, we can achieve $\sim$151\% energy savings at the cost of a $\sim$90\% increase in execution time. Furthermore, with the Euclidean distance metric, we achieve $\sim$200\% energy reduction at the expense of a $\sim$203\% runtime increase. The speedup-energy-delay provides settings that can yield higher runtime performance and lower energy savings over the Euclidean distance metric. These numbers are simple aggregations of per--GPU task energy and runtime values and represent ideal scenarios with no overhead. However, they indicate the power-capping impact on individual GPU tasks and how the whole application can potentially be affected, presenting opportunities for energy reduction. 
Although LSMS is a compute-intensive application, our results show that kernel-level power-capping can yield greater energy savings than application-wide tuning. By adapting power caps to individual GPU kernels based on their characteristics, we can significantly reduce energy consumption and enhance overall application efficiency.

GPU task-level power-capping strategies can employ these metrics and provide energy savings that approximate the observed impact or performance-energy patterns by focusing on efficient power-capping tuning and transitions for the runtime. Further optimization could combine these two metrics with user-defined goals. Energy savings could be prioritized over runtime costs, thereby enabling users to filter results based on acceptable performance trade-offs at the GPU task level.

%\section{Related Work}
%\input{src/related}
\section{Conclusions}
\begin{sloppypar}
This work demonstrates the effectiveness of applying dynamic, GPU task-specific power-capping strategies that use the NVIDIA GH200 superchip's automatic power-steering system. We evaluated two complementary metrics---speedup-energy-delay and Euclidean distance--based normalization---to determine optimal power limits per computational GPU task in the LSMS application. Comparing the optimal power-capping settings suggested by each one, the Euclidean distance--based normalization considers both runtime and energy equal optimization objectives, while affected by the data distribution differently than the speedup-energy-delay.
Our findings demonstrate significant energy savings achievable through fine-grained power management, thus emphasizing the balance between energy efficiency and performance. Future work will extend these methodologies, develop strategies for additional domains, and automate adaptive power-capping optimizations for easier integration into production HPC environments. These metrics can be seamlessly integrated into other GPU-accelerated HPC applications through widely adopted tools such as Score-P and PAPI. This integration makes them immediately beneficial to the broader HPC community for experimentation.
\end{sloppypar}

\begin{credits}
\subsubsection{\ackname} 
This material is based on work supported by the US Department of Energy's Office of Science, Advanced Scientific Computing Research program through EXPRESS: 2023 Exploratory Research for Extreme-Scale Science.
This research used resources of the Oak Ridge Leadership Computing Facility at Oak Ridge National Laboratory, which is supported by the Office of Science of the US Department of Energy under contract DE-AC05-00OR22725.

\subsubsection{\discintname}
The authors have no competing interests to declare that are
relevant to the content of this article. 
\end{credits}
\bibliographystyle{splncs04}
\bibliography{software.bib,apps.bib}

\begin{thebibliography}{10}
\providecommand{\url}[1]{\texttt{#1}}
\providecommand{\urlprefix}{URL }
\providecommand{\doi}[1]{https://doi.org/#1}

\bibitem{abdulsalam2015gpsup}
Abdulsalam, S., Zong, Z., Gu, Q., Qiu, M.: Using the greenup, powerup, and speedup metrics to evaluate software energy efficiency. In: 20Haidar,15 Sixth International Green and Sustainable Computing Conference (IGSC). pp.~1--8 (2015). \doi{10.1109/IGCC.2015.7393699}

\bibitem{10.1145/3581784.3607089}
Atchley, S., Zimmer, C., et~al.: Frontier: Exploring exascale. In: Proceedings of the International Conference for High Performance Computing, Networking, Storage and Analysis. SC '23, Association for Computing Machinery, New York, NY, USA (2023). \doi{10.1145/3581784.3607089}, \url{https://doi.org/10.1145/3581784.3607089}

\bibitem{azimi2023powercoord}
Azimi, R., Jing, C., Reda, S.: Powercoord: A coordinated power capping controller for multi-cpu/gpu servers. In: 2018 Ninth International Green and Sustainable Computing Conference (IGSC). pp.~1--9. IEEE (2018)

\bibitem{nvidia_power_thermals}
Corporation, N.: Nvidia grace performance tuning guide - power and thermals. \url{https://docs.nvidia.com/grace-perf-tuning-guide/power-thermals.html} (2024), accessed: 2025-02-28

\bibitem{10.1007/978-3-319-58667-0_21}
Eastep, J., Sylvester, S., Cantalupo, C., Geltz, B., Ardanaz, F., Al-Rawi, A., Livingston, K., Keceli, F., Maiterth, M., Jana, S.: Global extensible open power manager: A vehicle for hpc community collaboration on co-designed energy management solutions. In: Kunkel, J.M., Yokota, R., Balaji, P., Keyes, D. (eds.) High Performance Computing. pp. 394--412. Springer International Publishing, Cham (2017)

\bibitem{EISENBACH20172}
Eisenbach, M., Larkin, J., Lutjens, J., Rennich, S., Rogers, J.H.: Gpu acceleration of the locally selfconsistent multiple scattering code for first principles calculation of the ground state and statistical physics of materials. Computer Physics Communications  \textbf{211}, ~2--7 (2017), high Performance Computing for Advanced Modeling and Simulation of Materials

\bibitem{gerndt2016readex}
Gerndt, M.: The readex project for dynamic energy efficiency tuning. In: Proceedings of the ACM Workshop on Software Engineering Methods for Parallel and High Performance Applications. pp. 11--12 (2016)

\bibitem{haidar2019investigating}
Haidar, A., Jagode, H., Vaccaro, P., YarKhan, A., Tomov, S., Dongarra, J.: Investigating power capping toward energy-efficient scientific applications. Concurrency and Computation: Practice and Experience  \textbf{31}(6),  e4485 (2019)

\bibitem{10.1145/3703001.3724383}
Hernandez, O., Wang, T., Elwasif, W., Spiga, F., Tartaglione, F., Eisenbach, M., Miller, R.: Preliminary study on fine-grained power and energy measurements on grace hopper gh200 with open-source performance tools. In: Proceedings of the 2025 International Conference on High Performance Computing in Asia-Pacific Region Workshops. p. 11–22. HPC Asia '25 Workshops, Association for Computing Machinery, New York, NY, USA (2025). \doi{10.1145/3703001.3724383}, \url{https://doi.org/10.1145/3703001.3724383}

\bibitem{10.1007/978-3-642-31476-6_7}
Kn{\"u}pfer, A., R{\"o}ssel, C., Mey, D.a., Biersdorff, S., Diethelm, K., Eschweiler, D., Geimer, M., Gerndt, M., Lorenz, D., Malony, A., Nagel, W.E., Oleynik, Y., Philippen, P., Saviankou, P., Schmidl, D., Shende, S., Tsch{\"u}ter, R., Wagner, M., Wesarg, B., Wolf, F.: Score-p: A joint performance measurement run-time infrastructure for periscope,scalasca, tau, and vampir. In: Brunst, H., M{\"u}ller, M.S., Nagel, W.E., Resch, M.M. (eds.) Tools for High Performance Computing 2011. pp. 79--91. Springer Berlin Heidelberg, Berlin, Heidelberg (2012)

\bibitem{10.1007/978-3-030-43222-5_11}
Krzywaniak, A., Czarnul, P.: Performance/energy aware optimization of parallel applications on {GPU}s under power capping. In: Wyrzykowski, R., Deelman, E., Dongarra, J., Karczewski, K. (eds.) Parallel Processing and Applied Mathematics. Springer International Publishing (2020)

\bibitem{krzywaniak2022gpu}
Krzywaniak, A., Czarnul, P., Proficz, J.: Gpu power capping for energy-performance trade-offs in training of deep cnns. In: International Conference on Computational Science (ICCS). pp. 123--133. Springer (2022)

\bibitem{nvidia_energy_efficiency}
NVIDIA: Energy efficiency in high performance computing: Balancing speed and sustainability. \url{https://developer.nvidia.com/blog/energy-efficiency-in-high-performance-computing-balancing-speed-and-sustainability/} (2023), accessed: 2024-10-23

\bibitem{9610368}
Patrou, M., Kent, K.B., Siu, J., Dawson, M.: Energy and runtime performance optimization of node.js web requests. In: 2021 IEEE International Conference on Cloud Engineering (IC2E). pp. 71--82 (2021). \doi{10.1109/IC2E52221.2021.00021}

\bibitem{9969367}
Patrou, M., Kent, K.B., Siu, J., Dawson, M.: Optimizing energy efficiency of node.js applications with cpu dvfs awareness. In: 2022 IEEE 13th International Green and Sustainable Computing Conference (IGSC). pp.~1--8 (2022). \doi{10.1109/IGSC55832.2022.9969367}

\bibitem{saba2023ml}
Saba, I., Arima, E., Liu, D., Schulz, M.: Orchestrated co-scheduling, resource partitioning, and power capping on cpu-gpu heterogeneous systems via machine learning. In: Schulz, M., Trinitis, C., Papadopoulou, N., Pionteck, T. (eds.) Architecture of Computing Systems. pp. 51--67. Springer International Publishing, Cham (2022)

\bibitem{10.1145/3126908.3126945}
Sourouri, M., Raknes, E.B., Reissmann, N., Langguth, J., Hackenberg, D., Sch\"{o}ne, R., Kjeldsberg, P.G.: Towards fine-grained dynamic tuning of hpc applications on modern multi-core architectures. In: Proceedings of the International Conference for High Performance Computing, Networking, Storage and Analysis. SC '17, Association for Computing Machinery, New York, NY, USA (2017). \doi{10.1145/3126908.3126945}, \url{https://doi.org/10.1145/3126908.3126945}

\bibitem{10820573}
Terboven, C., Liem, R., Gracia, J., Haldar, K., Engels, J., Giesselmann, P., Brayford, D., Wilde, T., Simmendinger, C., Marquardt, M., Eitzinger, J., Gruber, T.: Ee-hpc a framework for energy efficient hpc system management. In: SC24-W: Workshops of the International Conference for High Performance Computing, Networking, Storage and Analysis. pp. 1878--1882 (2024). \doi{10.1109/SCW63240.2024.00236}

\bibitem{zhao2023sustainable}
Zhao, D., Samsi, S., McDonald, J., Li, B., Bestor, D., Jones, M., Tiwari, D., Gadepally, V.: {Sustainable supercomputing for AI: GPU power capping at HPC scale}. In: Proceedings of the 2023 ACM Symposium on Cloud Computing. pp. 588--596 (2023)

\end{thebibliography}

\end{document}